\documentclass[journal=jcisd8,layout=twocolumn,manuscript=article]{achemso}

\usepackage[T1]{fontenc}
\usepackage{graphicx}

\usepackage{amsmath}
\usepackage{times}
\usepackage{soul}
\usepackage[normalem]{ulem}
\usepackage[version=3]{mhchem}
\usepackage{dsfont}

\newcommand{\rb}{\mathbf{r}}

\newcommand{\Fb}{\mathbf{F}}
\newcommand{\Wb}{\mathbf{W}}

\newcommand{\dr}{\text{d} \rb}

\newcommand{\icm}{cm$^{-1}$}

\newcommand{\etal}{\textit{et al}.\ }
\newcommand{\ie}{\textit{i}.\textit{e}.}

\newcommand{\hili}{}
\newcommand{\remove}[1]{}

\usepackage{xcolor}

\title{Raman spectra of amino acids and peptides from machine learning polarizabilities}

\author{Ethan Berger}
\affiliation{Microelectronics Research Unit, Faculty of Information Technology and Electrical Engineering, University of Oulu, P.O. Box 4500, Oulu, FIN-90014, Finland}
\author{Juha Niemelä}
\affiliation{Faculty of Biochemistry and Molecular Medicine, University of Oulu, Oulu, Finland}
\author{Outi Lampela}
\affiliation{Biocenter Oulu and Faculty of Biochemistry and Molecular Medicine, University of Oulu, Oulu, Finland}
\author{André H. Juffer}
\affiliation{Biocenter Oulu and Faculty of Biochemistry and Molecular Medicine, University of Oulu, Oulu, Finland}
\author{Hannu-Pekka Komsa}
\affiliation{Microelectronics Research Unit, Faculty of Information Technology and Electrical Engineering, University of Oulu, P.O. Box 4500, Oulu, FIN-90014, Finland}
\email{hannu-pekka.komsa@oulu.fi}

\begin{document}
\begin{abstract}
    Raman spectroscopy is an important tool in the study of vibrational properties and composition of molecules, peptides and even proteins. Raman spectra can be simulated based on the change of the electronic polarizability with vibrations, which can nowadays be efficiently obtained via machine learning models trained on first-principles data. 
    However, the transferability of the models trained on small molecules to larger structures is unclear and direct training on large structures in prohibitively expensive.
    In this work, we first train two machine learning models to predict polarizabilities of all 20 amino acids. Both models are carefully benchmarked and compared to DFT calculations, with neural network method found to offer better transferability. By combining machine learning models with classical force field molecular dynamics, Raman spectra of all amino acids are also obtained and investigated, showing good agreement with experiments. The models are further extended to small peptides. We find that adding structures containing peptide bonds to the training set greatly improves predictions even for peptides not included in training sets.
\end{abstract}

\section{Introduction}

Raman spectroscopy is an important tool in modern research, as it represents a non-destructive method to study the vibrational properties of both solids and molecules. In biochemistry, it can be used to study the complex structures of large molecules such as proteins or peptides \cite{Kuhar2018,Kuhar2021}. For instance, Raman spectra contain important information on the folding of proteins \cite{Oladepo2011,Serrano2012}. In this case, peaks pertaining to the peptide bonds (typically amide I and II) can be used to understand the impact of folding on the Raman spectra \cite{Kuhar2020,Xu2007}. Another highly promising development is the use of surface-enhanced Raman spectroscopy to study the primary structures of protein, \ie\ their sequence of amino acids. As the protein moves across the plasmonic hotspot, 
this technique could allow one to record sequentially the spectra of individual amino acids \cite{Huang2020,Zhao2022}. In all these cases, however, it is still necessary to have good reference spectra in order to perform correct assignment. While spectra of biomolecules \cite{DeGelder2007} and amino acids \cite{Jenkins2005,Zhu2011} have been investigated many times, their interpretation remains challenging. In this context, simulation of Raman spectra could produce such reference data and further help understanding the experimental observations.  

Simulations of Raman spectra usually rely on calculating harmonic vibrational modes and Raman tensors \cite{Loudon2001,Porezag96_PRB,Bagheri2023}. For large molecules such as peptides, calculations have to be repeated many times for each possible conformations, making it highly tedious and computationally expensive. 
\hili{\remove{By using molecular dynamics (MD), Raman spectra is naturally obtained for every visited conformations, making the analysis much easier.}
When using molecular dynamics (MD), all visited conformations are included in the trajectory. The resulting Raman spectra therefore represent an average over these conformations, making the analysis much easier.} 
Raman spectra from MD is obtained using the Fourier transform of the polarizability autocorrelation function \cite{BernePecora,Cardona1982,Putrino02_PRL,Thomas13_PCCP}. High quality Raman spectra therefore rely on producing accurate MD trajectories as well as correctly predicting polarizabilities along these trajectories.
First-principle calculations could be used to obtain both trajectories and polarizabilities, but the computational cost increases quickly with system size. Such methods are limited to small systems and not applicable to peptides or proteins. For this kind of molecules, trajectories can be obtained from classical force-fields. These represent good compromise between accuracy and speed, as they can routinely produce MD trajectories of hundreds of nanoseconds for large systems containing thousands of atoms. On the other hand, no clear candidates exist for the prediction of polarizabilities. 

Bond polarizability models (BPM) have been used to predict polarizabilities of solids and molecules \cite{Umari2001_PRB,paul2023,Berger2023_2}. This type of models relies on the bond lengths between atoms and one typical application has been prediction of polarizabilities of alkanes, which only contain C-C and C-H bonds \cite{Smirnov2006,Chen2017}. This makes their modeling relatively simple and it was shown that the model is also applicable to larger molecules than those included during the training step. However, application of BPMs to amino acids would be challenging due to the various kind of bonds present in these molecules.
Another model for polarizabilities is the Thole model \cite{Thole1981,Duijnen1998}. It is based on the dipole interaction and takes atomic polarizabilities as parameters. While atomic polarizabilities are independent of the atom position, changes in atomic positions are accounted through the dipole field tensor. In its initial form, this tensor could lead to infinite polarization \cite{Applequist1972}. To overcome this problem, Thole proposed various continuous shapes \cite{Thole1981,Wang2011}, which all relied on a screening length as parameters. This length and the atomic polarizabilities therefore represent parameters to be optimized, which has for example been done for amino acids using first-principles calculations \cite{Swart2004}.
Alternatively, machine learning (ML) methods can now efficiently predict tensorial properties such as polarizabilities and could easily be applied to peptides or larger molecules. There exist many types of machine learning algorithms, but not all of them have been extended to predict tensorial properties. 
Neural networks (NN) were among the first to be applied for energy prediction \cite{Behler_2007}. This type of ML model are now commonly used to produce efficient force fields for MD simulations \cite{Behler2021,Fan2021} and has also been applied for the prediction of polarizabilities and Raman spectroscopy \cite{Sommers2020,Shang2021,Han2022}. Feng \etal combined dipole-interaction model similar to the Thole model and NN to learn the atomic polarizabilities \cite{Feng2023}. \hili{Other examples of NN applicable to tensorial quantities are TNEP \cite{Xu2023}, e3nn \cite{geiger2022e3nn,Schienbein2023} or FieldSchNet \cite{Gastegger2021}.}
Another ML method is Gaussian process regressor (GPR), which uses kernel functions to describe the similarity between atomic configurations \cite{Bartok_2010}. Similarly to NN, GPR have been applied to obtain force fields \cite{Jinnouchi_2019_1,Jinnouchi_2019_2,Deringer2021} as well as polarizabilities and Raman spectra \cite{Grisafi_2018,Raimbault2019,Wilkins_2019,Zauchner_2021,Grumet2023}. \hili{For example, GPR has been combined with path-integral MD trajectories to obtain the vibrational spectra of water \cite{Kapil2020,Shepherd2021}.} More recently, GPR was also extended for prediction of electronic densities and its response to electric field, which consequently allows one to obtain polarizabilities \cite{Lewis2021,Lewis2023,Grisafi2023}. 
One remaining obstacle to ML-based methods is transferability, \ie, the reliability of the predictions when the model is applied to molecules outside its training set. A model trained solely on paracetamol molecules can reasonably predict polarizabilities and Raman spectra of paracetamol crystals \cite{Raimbault2019} and BPM and GPR trained for alkanes of different sizes can also successfully predict the Raman spectra of large molecules
\cite{Fang2023}.
In the case of amino acids, it has been demonstrated that a single ML model trained on small molecules (less than 7 heavy atoms) can also predict total molecular polarizabilities for larger molecules (typically single amino acids) \cite{Wilkins_2019}. However, to the best of our knowledge, application of such ML models for simulating the Raman spectra of amino acids (let alone peptides) has not been demonstrated. 
Since atomic structures of amino acids, and consequently of peptides, are all similar, develop a single ML model for them should be feasible.

In this paper, we train and compare two ML models based on NN and GPR for the prediction of polarizabilities and Raman spectra. Each model is first trained only on amino acids and the resulting models are then extended to small glycine-based peptides. Resulting final models are used to simulate the Raman spectra, which are compared to those obtained from explicit DFT polarizabilities as well as the Thole model for the glycine molecule. Spectra of all 20 amino acids are then compared with experimental results from the literature. Finally, the Raman spectra of met- and leu-enkephalins are investigated. In this case, spectra are again compared to results from the Thole model and committee error estimates are also used to assess the model accuracy.

\section{Methods}

Raman spectra $I(\omega)$ can be obtained from the Fourier transform of the polarizability auto-correlation function \cite{BernePecora,Cardona1982,Putrino02_PRL,Thomas13_PCCP}. In the case of molecules, both the isotropic and anisotropic part of the spectra can be obtained. Throughout this work, we focus on the isotropic part $I_\text{iso}$ which can be written as
\begin{equation}\hili{
I_\text{iso}(\omega) \propto \int \langle \bar{\chi}(\tau) \bar{\chi}(t+\tau)\rangle_\tau \ e^{-i\omega t} dt }
\label{equ:RamanMD}
\end{equation} 
where $\bar{\chi}(t)=\frac{1}{3}\left(\chi_{xx}+\chi_{yy}+\chi_{zz}\right)$ denotes the isotropic average of the polarizability tensor at time $t$, depending on the positions of atoms at that time, and $\langle\bar{\chi}(\tau) \bar{\chi}(t+\tau)\rangle_\tau$ is the polarizability auto-correlation function. Thus, in addition to the MD trajectory, also the polarizabilities along this trajectory need to be evaluated. Here classical force fields are used to perform MD and ML models are then used to predict polarizabilities. The rest of this section introduces the polarizability models as well as computational details regarding MD and DFT calculations.

\subsection{Gaussian process regressor} 

Symmetry-adapted Gaussian process regressor (SA-GPR) is an extension of GPR to predict tensorial properties \cite{Grisafi_2018}. It has been successfully applied to molecules and periodic systems to predict their polarizabilities \cite{Grisafi_2018,Raimbault2019,Wilkins_2019,Berger2023_2}. In general GPR, the quantity of interest $y$ is obtained using a kernel function $k(X,X')$, which describes the similarity between atomic configurations $X$ and $X'$. This reads
\begin{equation}
    y(X) = \sum_{n}\omega_{n}k(X,X_n),
    \label{equ:GPR}
\end{equation}
where $\omega_n$ are weights and the index $n$ runs over a reference training set. By considering all polarizabilities in the training set, equation \ref{equ:GPR} can be written as a system of equations $y=\omega k$ and solved to find the weights $\omega$. Once these are known, equation \ref{equ:GPR} can be used to predict polarizabilities of new structures. In SA-GPR, the kernel is replaced by a tensorial kernel $K(X,X')$ \cite{Grisafi_2018}. For large systems, it is also beneficial to consider only local environment $X^i$, which are usually obtained by considering atoms within a cutoff distance around atom $i$. Polarizabilities $\chi$ then reads
\begin{equation}
    \chi = \frac{1}{N_iN_j}\sum_{i,j,n}\omega_{n}K(X^i,X^j_n)
    \label{equ:SAGPR}
\end{equation}
where $N_i$ and $N_j$ are the number of local environments in structures $X$ and $X_n$, respectively. One popular choice for the kernel is smooth overlap of atomic positions (SOAP) \cite{Bartok_2013}, which has already been used in many applications \cite{Szlachta2014,De2016,Deringer2017,Fujikake2018,Fabrizio2019,Veit2020} and is also used in this work. The tensorial kernel then takes the form
\begin{equation}
    K(X,X') = \int \text{d}\hat{R}\ \mathbf{D}(\hat{R})\left|\int\dr\ \rho(\rb)\rho'(\hat{R}\rb)\right|^2,
    \label{equ:TEN_kernel}
\end{equation}
where $\mathbf{D}$ are the Wigner D-matrices and $\rho$ and $\rho'$ are the atomic densities of configuration $X$ and $X'$, respectively. By expanding the atomic densities using radial functions $g_n(r)$ and spherical harmonic angular functions $Y_{lm}(\hat{r})$, the kernel in equation \ref{equ:TEN_kernel} can be rewritten in terms of power spectra \cite{Darby2022,Grisafi_2018}. The number of radial and angular functions considered in the expansion have to be chosen, as well as the cutoff radius used for the local environments.

\subsection{Neuroevolution potential} 

Neuroevolution potential (NEP) is a feed-forward neural network with a single hidden layer \cite{Fan2021,Fan2022a,Fan2022b}. One specificity of NEP is that it uses the separable natural evolution strategy \cite{Fan2021,Schaul2011} to optimize the NN parameters. Similar to SA-GPR, local environment around atom $i$ is transformed into a descriptor array $q_\nu^i$ using an expansion over radial and angular functions (detailed definition of NEP descriptors can be found in Ref.\ \citenum{Fan2022b}). These are used as the input layer and are fed to the hidden layer using an activation function which takes the form
\begin{equation}
    x_\mu^i = \tanh\left(\sum_\nu \omega_{\nu\mu}^{(1)} q_\nu^i+b_\mu^{(1)}\right),
    \label{equ:NEP_hidden}
\end{equation}
where $\omega_{\nu\mu}^{(1)}$ are the connection weights, $b_{\mu}^{(1)}$ are the hidden layer bias and hyperbolic tangent is used as activation function. The hidden layer state vector $x_\mu^i$ is further used to obtain the local energy $U^i$ as
\begin{equation}
    U^i = \sum_\mu \omega_\mu^{(2)}x_\mu^i+b^{(2)},
    \label{equ:NEP_out}
\end{equation}
where $\omega_\mu^{(2)}$ are connection weights between the hidden layer and the output and $b^{(2)}$ is the output bias. Total energy $U$ of the system is then obtained as the sum of local energies $U=\sum_i U^i$. In NEP, tensorial properties (usually the virial $\Wb$) are also computed for local environment. For local virials $\Wb^i$, it takes the form
\begin{equation}
    \Wb^i = \sum_{j\neq i}\rb_{ij}\otimes\frac{\partial U^j}{\partial\rb_{ji}}
    \label{equ:NEP_virial}
\end{equation}
where $\rb_{ij}$ is the vector between atoms $i$ and $j$ and $\otimes$ denotes the outer product between the two vectors. Given the form of $U^i$ in equations \ref{equ:NEP_hidden} and \ref{equ:NEP_out}, derivatives $\frac{\partial U_j}{\partial\rb_{ji}}$ can be obtained analytically, making it easy to compute. For force fields, parameters $\omega_{\nu\mu}^{(1)}$, $\omega_\mu^{(2)}$, $b_{\mu}^{(1)}$ and $b^{(2)}$ (as well as parameters used in descriptors) are then optimized to minimize the error on energy, forces and virial.

The NEP formalism was recently extended to predict tensorial properties (in which case it is refer\hili{r}ed to as TNEP) \cite{Xu2023}. While not being related to the energy in the same way, polarizability tensors have the same symmetry properties \hili{as the} virials. Using this fact, polarizabilities can be obtained in the same way as virials. For local polarizabilities $\chi^i$, equation \ref{equ:NEP_virial} reads 
\begin{equation}
    \chi^i = Q^i \cdot \mathds{1} - \sum_{j\neq i}\rb_{ij}\otimes\frac{\partial Q^j}{\partial\rb_{ji}},
    \label{equ:NEP_pol}
\end{equation}
where $\mathds{1}$ denotes the identity matrix and $Q^i$ are obtained through the NN (equations \ref{equ:NEP_hidden} and \ref{equ:NEP_out}). Since \hili{$Q^i$} do not represent any physical quantities, they are not used during optimization. Parameters are therefore optimized to only minimize errors on polarizabilities \hili{$\chi^i$}.

\subsection{Thole} 

The Thole model is an empirical model based on the dipole-dipole interaction \cite{Thole1981,Duijnen1998}. For a collection of atoms placed in an homogeneous electic field $\Fb$, the dipole moment $\mu_p$ of atom $p$ can be written as \cite{Applequist1972}
\begin{equation}
    \mu_p = \alpha_p \left[ \Fb_p - \sum_{q\neq p}T_{qp}\mu_q\right]
\end{equation}
where $\alpha_p$ is the atomic polarizability of atom $p$ and $T_{qp}$ is the dipole field tensor. This equation can be seen as a system of equations of the form $\mu = A\Fb$, with matrix $A$ being defined as
\begin{equation}
    A = \left[\alpha^{-1} + T\right]^{-1},
\end{equation}
where matrix $\alpha$ contains the atomic polarizabilities as diagonal and $T$ is formed by matrices $T_{qp}$ (with zeros in the diagonal). In this way, the usual form for the relation between dipolar moment and polarizability is recovered, albeit using an effective polarizability $A$. It can be also noted that in the case $T=0$ (\ie\ the dipole-dipole interaction is neglected) $A$ is simply given by the atomic polarizabilities.
The molecular polarizability is then retrieved by summing the elements of $A$
\begin{equation}
    \chi_{ij} = \sum_{qp} (A_{ij})_{qp}.
\end{equation}

Thole proposed various shapes for the dipole field tensors $T$ which depend on a screening length $a$ \cite{Thole1981}. In this work, we use the exponential shape and the parameters (length $a$ and atomic polarizabilities) optimized for amino acids presented in Ref.\ \citenum{Swart2004}. 

\subsection{Computational details} 

The MD simulations were performed using GROMACS version 2022.4 \cite{VanderSpoel2005} with the CHARMM27 force field \cite{Bjelkmar2010} and the TIP3P water model \cite{Jorgensen1983}.

The initial structures for met-enkephalin and leu-enkephalin were extracted from PDB entries 5E33 and 5E3A (from RCSB.org). For the single amino acids and the glycine based small peptides, the initial structures were created by using the 'build' option in PyMOL 2.5.0 Open-Source \cite{PyMOL}. The protonation states were set according to p$K_a$ values assuming neutral pH. For each simulation, the molecule of interest was solvated in water in a cubic simulation box with a 1 nm distance between the molecule and the edges of the box. For molecules which had a net charge, the total charge of the simulation box was set to zero by the addition of Na$^+$ or Cl$^-$ ions.

Before the production MD runs, an energy minimization run and NVT and NPT equilibrations were carried out. The NVT and NPT equilibrations were run for 100 ps with a time step of 2 fs. For the production runs, the time step was 0.5 fs and simulation length was 100 ns. The atomic coordinates were stored every fifth step (every 2.5 fs). The reason for the unusually short time step in the production runs was that it was necessary to capture the high frequency hydrogen vibrations. The short-range electrostatic cutoff and short-range van der Waals cutoff were both 1.0 nm. Particle Mesh Ewald \cite{Darden1993} was used for long-range electrostatics. \hili{Production runs were performed in the NPT ensemble.} V-rescale \cite{Bussi2007} was used as the temperature coupling method, and Parrinello-Rahman \cite{Parrinello1981,Nose1982} was used as the pressure coupling method. The reference temperature for temperature coupling was 300 K, and the reference pressure for pressure coupling was 1 bar.

DFT calculations are performed using the cp2k software \cite{VandeVondele2005,Kuhne2020}. \hili{Note that while all classical MD simulations are performed in presence of water, all DFT calculations are done in vacuum as the effect of solvation on the polarizability is found to be very small (see Fig.\ S1 in the SI for details)}. The B3LYP functional \cite{Stephens1994} is used with GTH pseudopotentials \cite{Goedecker1996,Hartwigsen1998} and double zeta Gaussian basis set \cite{VandeVondele2007}. A cutoff energy of 800 Ry is set for the expansion of the electron density in plane waves. Polarizabilities are obtained from linear response, as directly implemented in cp2k \cite{Putrino02_PRL}. \hili{For glycine, polarizabilities are calculated for the first $\sim$9 ps of the trajectory, which allow to obtain Raman spectra at the DFT level. This set of polarizabilities is also used to train ML models for glycine only. 3000 structures are kept for training and the remaining 653 are used for validation. For every amino acid, polarizabilities are calculated for 200 equally separated structures from productions runs. Training sets are created by randomly taking 150 structures for each system (3150 structures in total), while the remaining 50 are used for validation (1050 in total). For peptides, the exact same procedure is done for GG and GGG, while for GAG and GSG 200 structures are used for validations. This lead to a training and validation set containing 3450 and 1550 structures, respectively. In the end, this leads to three different data sets: one for glycine only, one for amino acids only and one for amino acids and peptides.}

Hyperparameters for TNEP and SA-GPR are chosen to be similar for a more direct comparison. For both methods, local environment are obtained using a cutoffs of 4 \AA\ and 4 radial and angular functions are used \hili{to build} descriptors. Additionally, 30 neurons are used in the hidden layer of TNEP, which corresponds to the recommended value.

\section{Results}
\subsection{Training of machine learning models}
 
\begin{figure}[t] 
    \centering
    \includegraphics[width=3.33in]{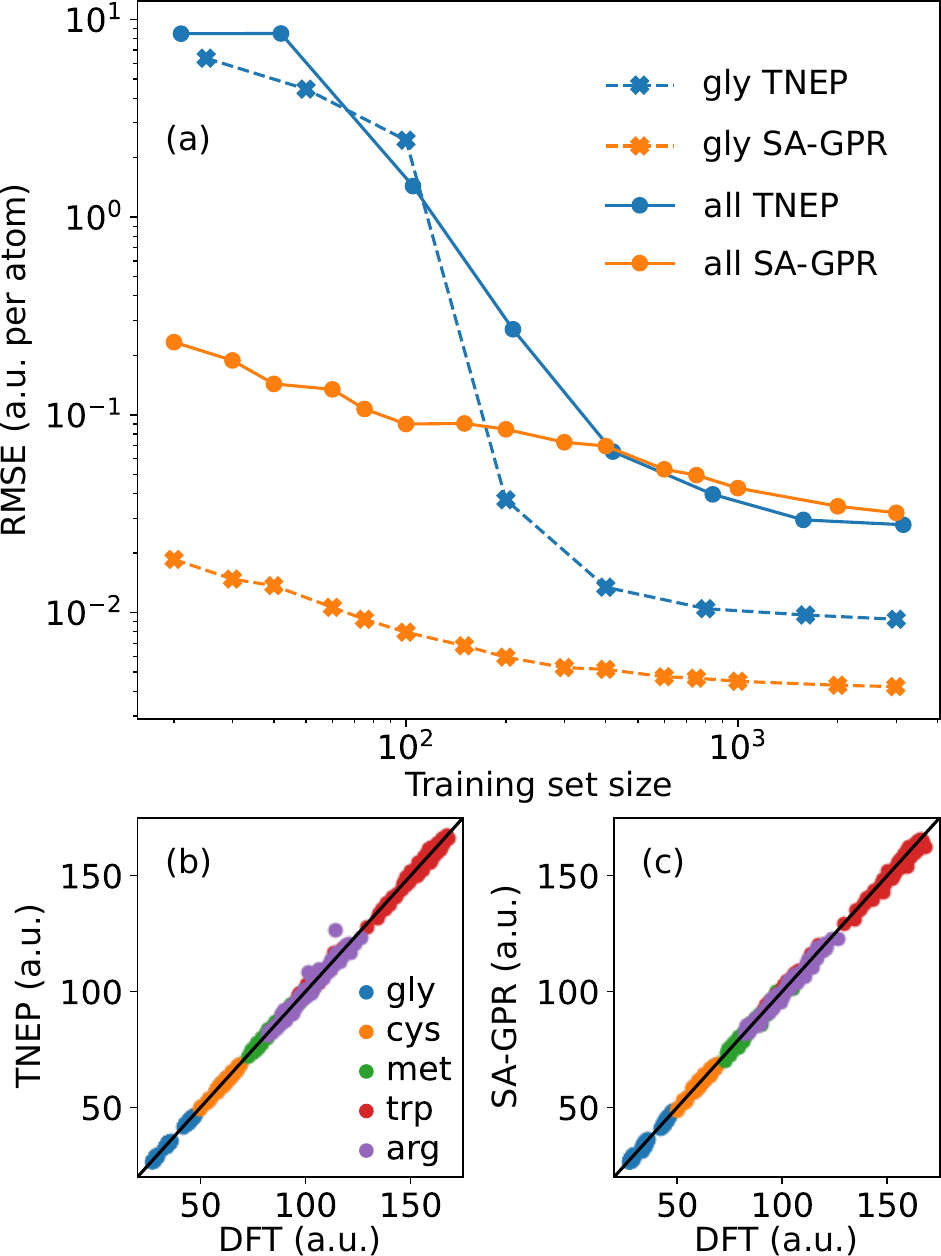}
    \caption{(a) Convergence of the \hili{polarizability RMSE of the validation set} with the training set size. Results for models trained on all amino acids (dots with solid lines) and on only glycine (crosses with dashed lines) are both represented. Additionally, TNEP (in blue) and GPR (in orange) are compared. (b)-(c) Comparison between polarizabilities from DFT and predictions from TNEP and GPR, respectively. The models used in (b) and (c) correspond to the final points in panel (a).}
    \label{fig:Fig1}
\end{figure}

Models are first trained only for glycine, which is the simplest amino acid. Fig.\ \ref{fig:Fig1}(a) shows the evolution of the root mean squared error (RMSE) per atom with increasing training set size and results from TNEP and SA-GPR are compared. RMSE is calculated from a validation set kept constant even when increasing the size of the training set. 
For a single molecule, SA-GPR is found to be more accurate than TNEP, with respective RMSE of $4.21\cdot 10^{-3}$ and $9.24\cdot 10^{-3}$ au per atoms when using 3000 structures for training. For smaller training set sizes, SA-GPR accuracy is much better than TNEP. \hili{Also note that the training set of glycine is very dense, with structures being separated by only 2.5 fs. SA-GPR seems to perform better for such dense training sets.} Models are then extended to all 20 amino acids and results are also presented in Fig.\ \ref{fig:Fig1}(a). Similarly to glycine models, SA-GPR is more accurate than TNEP for small training sets, even though addition of different kind of molecules clearly lowered the accuracy. For larger training sets containing more than 500 structures, SA-GPR and TNEP lead to similar accuracy, with RMSE of $3.20\cdot 10^{-2}$ and $2.78\cdot 10^{-2}$ au per atoms, respectively. Fig.\ \ref{fig:Fig1}(b) and (c) show predicted polarizabilities of the test set from TNEP and SA-GPR models. Predictions are shown for a few distinct amino acids and both models predict polarizabilities accurately for all of them, whether it is small (glycine), contains sulfur atom (cysteine and methionine), contains an aromatic ring (tryptophan), or is charged (arginine). 

\begin{figure}[t] 
    \centering
    \includegraphics[width=3.33in]{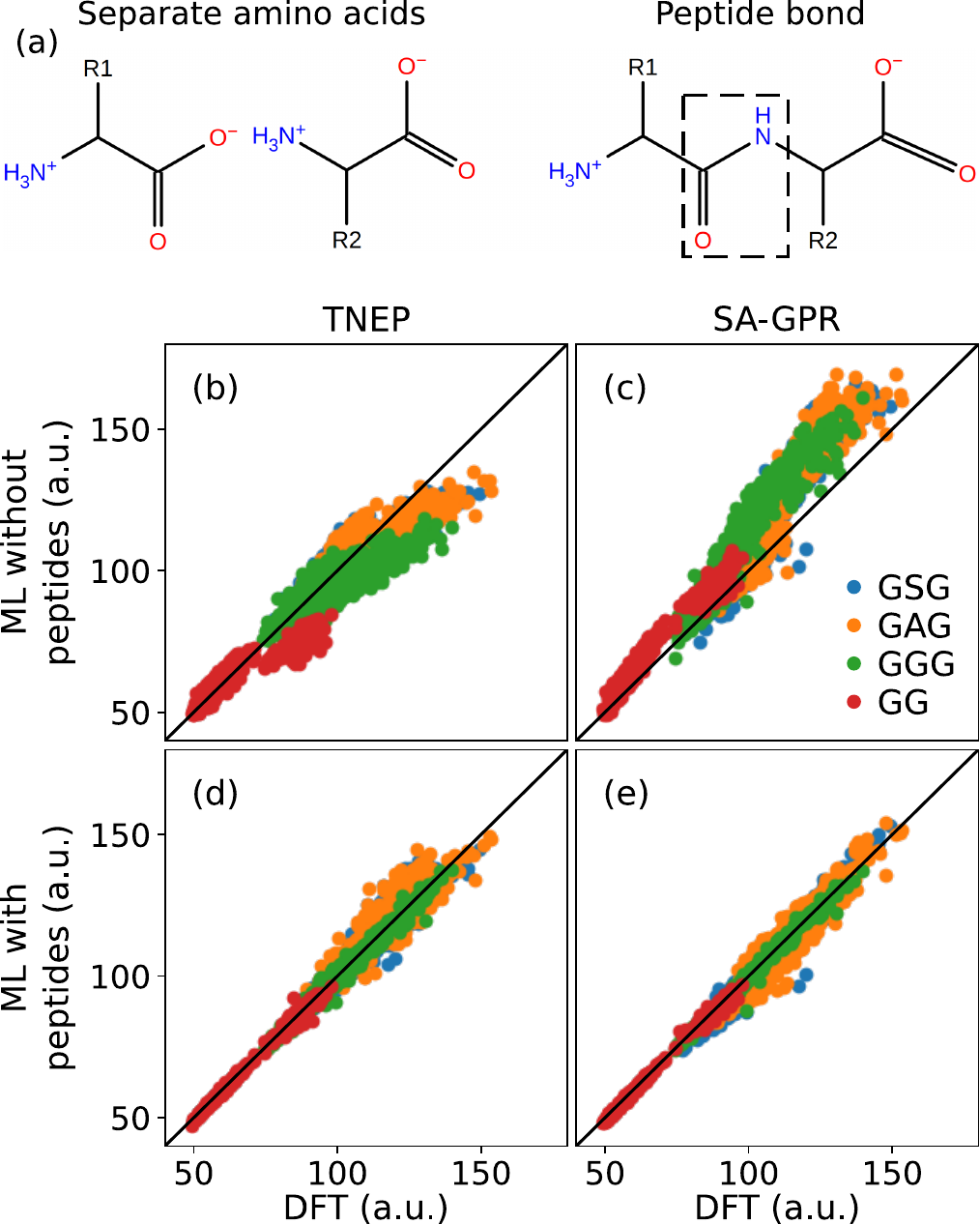}
    \caption{(a) Schematic showing the formation of peptide bonds. Comparison of polarizabilities between DFT and ML models for glycine-based peptides: (b) TNEP and (c) SA-GPR without peptides in the training set, and (d) TNEP and (e) SA-GPR including GG and GGG during training.}
    \label{fig:Fig2}
\end{figure}

Models are then applied to peptides, which are small chains of amino acids. One of the oxygen from the \ce{COO-} group of an amino acid and two hydrogens from \ce{NH3} of another amino acids might form water, leading to covalent bonding between the two amino acids called peptide bond. A schematic showing the principle behind peptide bonding is shown in Fig.\ \ref{fig:Fig2}(a). This lead to new N-C-O environment (called amide). 
While models have been trained for isolated amino acids so far, their application to peptides requires strict testing, as local environment of the peptide bonds has not been learned. Therefore, both models are tested on simple peptides, namely glycine-glycine (GG), glycine-glycine-glycine (GGG), glycine-alanine-glycine (GAG) and  glycine-serine-glycine (GSG). Fig.\ \ref{fig:Fig2}(b) and (c) compare polarizabilities from DFT with those from TNEP and SA-GPR models, respectively. Models used here are \hili{\remove{ those previously presented (see Fig.\ \ref{fig:Fig1}) and}} trained using only amino acids. Both models lead to inaccurate polarizabilities. \hili{\remove{SA-GPR constantly overestimates polarizabilities, leading to a constant shift}}. Such differences might appear because the training set do not contain any structures with peptide bonds. New models containing di- and tri-glycine are therefore trained and their resulting predictions are presented in Fig.\ \ref{fig:Fig2}(d) and (e). Expectedly, both models now lead to very accurate polarizabilities for GG and GGG\hili{, with RMSE of $5.79\cdot10^{-2}$ and $6.96\cdot10^{-2}$ au per atoms for TNEP and $5.34\cdot10^{-2}$ and $4.79\cdot10^{-2}$ au per atoms for SA-GPR, respectively. These values compare well with those previously found for amino acids}. TNEP also shows satisfactory accuracy for other peptides GAG and GSG\hili{, with RMSE of $1.61\cdot10^{-1}$ and $1.69\cdot10^{-1}$ au per atoms, respectively. \remove{Results from SA-GPR are however worse as part of the shift previously observed still remains in the new model.} For SA-GPR, we find that increasing the radial cutoff to 6 \AA\ leads to more accurate models (see Fig.\ S2 in the SI). This value is therefore used for models represented in Fig.\ \ref{fig:Fig2}(c) and (e) as well as applications to Raman spectra. With this improvement, SA-GPR gives RMSE of $1.44\cdot10^{-1}$ and $1.62\cdot10^{-1}$ au per atoms for GAG and GSG, respectively.} \hili{\remove{While both methods lead to accurate predictions for amino acids, TNEP performs better for peptides not included in the training set. On the other hand, SA-GPR is found to perform better when trained for a single molecule.}} This shows the importance of including peptide bonds in the training set. The accuracy of both models is satisfactory for single amino acids and short peptides, allowing us to apply them to MD trajectories to investigate Raman spectra.

For larger peptides, polarizabilities at the DFT level are too expensive to be computed for many structures and it becomes crucial to \hili{estimate the errors pertaining to the ML model.} In this work, committee error estimates \hili{(CEE)} are used to \hili{evaluate} such errors \cite{Schran2020,Carrete2023}. 10 TNEP models are trained using the same training set. Given the stochastic nature of the TNEP training, each model will have different optimized parameters and therefore resulting in slightly different predictions. 
The quality of the model predictions can be assessed by comparing the model predictions, which is here quantified by evaluating the standard error between polarizabilities predicted by the committee models. Similarly, the spectra obtained using each committee model can be compared to assess their quality. More information regarding committee error estimates can be found in the SI (see in particular Figs.\ S3 and S4). \hili{It is important to note that in our case CEE always underestimates the RMSE and is therefore only an estimation. Additionally, CEE only accounts for the epistemic uncertainty \cite{Thaler2023}, while other sources of errors such as DFT approximations, accuracy of classical force fields and limited MD sampling are not included.}


\subsection{Raman spectra of single amino acids}

\begin{figure*}[t] 
    \centering
    \includegraphics[width=7in]{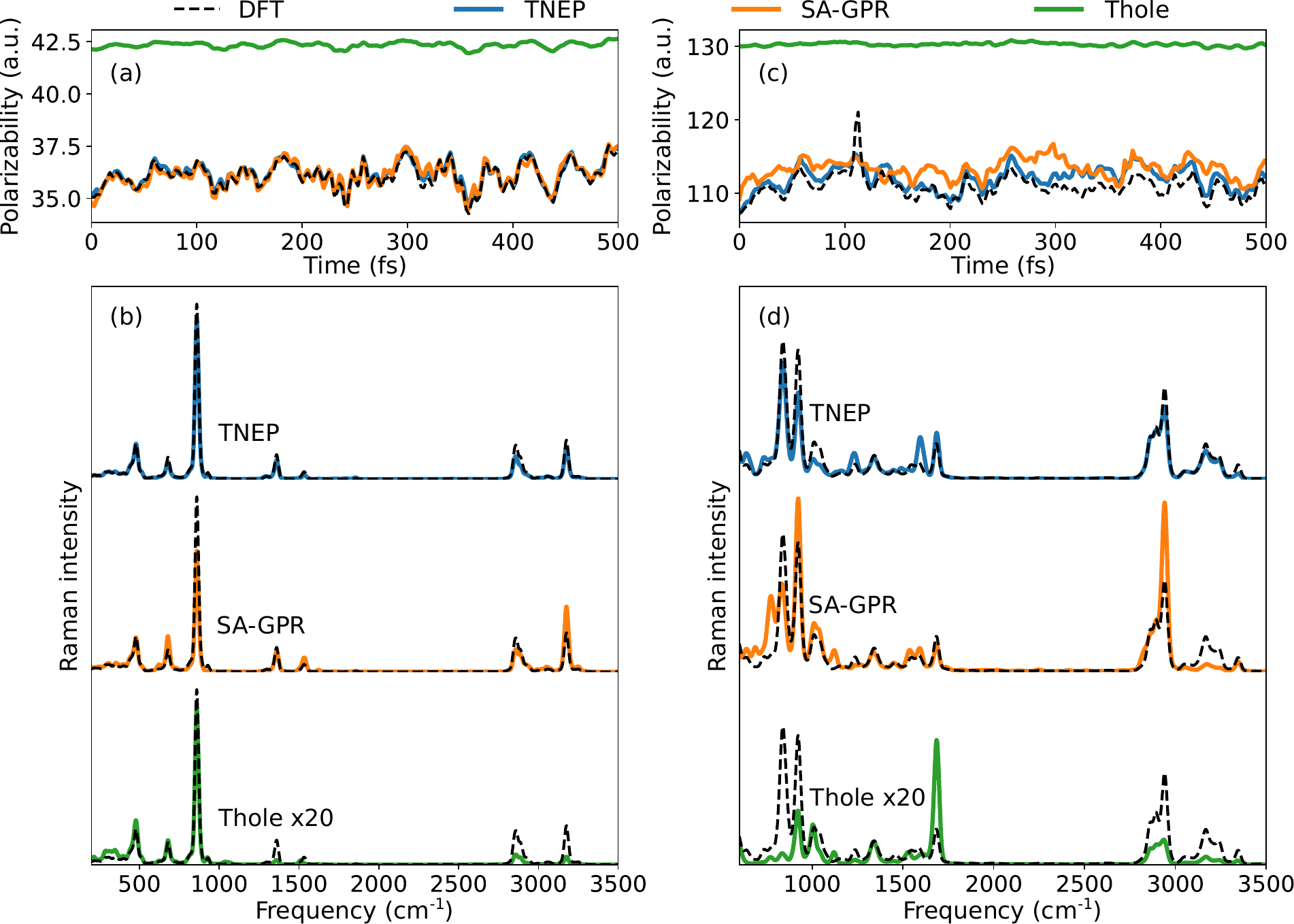}
    \caption{(a) Isotropic polarizability with time and (b) the resulting Raman spectra of glycine. (c-d) Isotropic polarizability with time and Raman spectra of the GAG peptide. DFT results are compared with TNEP, GPR and Thole model using the same line color in all panels.}
    \label{fig:Fig3}
\end{figure*}

\begin{figure}[t] 
    \centering
    \includegraphics[width=3.33in]{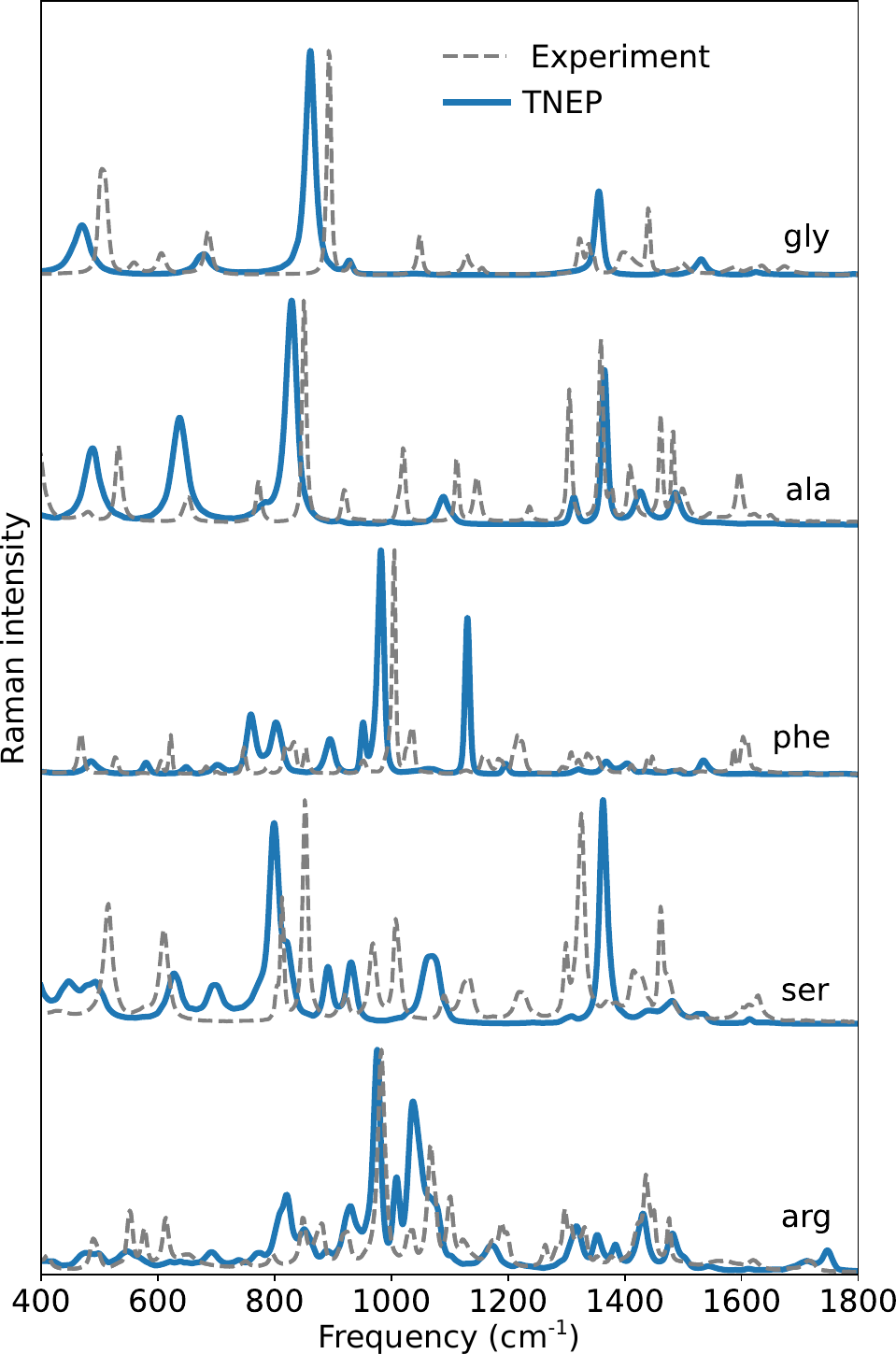}
    \caption{Comparison with experimental results of Raman spectra for a few amino acids. Experimental results are taken from Ref.\ \citenum{DeGelder2007}.}
    \label{fig:Fig4}
\end{figure}

Raman spectra are obtained by combining classical MD trajectories with polarizabilities from ML models. \hili{\remove{Before applying it to all amino acids and peptides, t}T}he quality of Raman spectra from each model is \hili{first} tested \hili{\remove{for glycine}} by comparing them with \hili{results from DFT calculations} Results are presented in Fig.\ \ref{fig:Fig3}. \hili{In particular, Fig.\ \ref{fig:Fig3}(a) shows the polarizability with time for glycine and Raman spectra are then compared in Fig.\ \ref{fig:Fig3}(b).} Fluctuations of the DFT polarizabilities are well reproduced by both ML models. The Thole model compares much worse, which is not surprising given that (i) it is not trained on the same DFT data, (ii) it is intended to reproduce polarizabilities of molecules but not the polarizability fluctuations with vibrations, and (iii) the structure of Thole model is not well suited for describing polarizability changes with vibrations as there is only a weak dipolar coupling between atoms (and thus fluctuations are underestimated).
Raman spectra are then shown in Fig.\ \ref{fig:Fig3}(b). Peak positions mostly depend on the trajectories and hence the discrepancies between the three models and DFT are small. On the other hand, intensities are more sensitive to the choice of polarizability model. TNEP is found to be in great agreement with DFT results for all peaks. On the other hand, SA-GPR shows less accurate intensities, with the peak at 860 \icm\ being underestimated and peaks around 3000 \icm\ being overestimated. \hili{\remove{Additionally, SA-GPR spectra is quite noisy between 1600 and 2700 \icm\ (shown in insets) where DFT spectrum shows no peaks.}} Finally, spectrum from Thole model shows greatly underestimated intensities [multiplied by a factor 20 in Fig.\ \ref{fig:Fig3}(b)], but after normalization the main features are still reasonably well reproduced except for the underestimation of peak intensities around 3000 \icm. 
\hili{Additionally, we test the accuracy of Raman spectra for different training set sizes (see Fig.\ S5 in the SI). Raman spectra from SA-GPR are found to be more accurate than TNEP for small training sets, in good agreement with what was previously observed in Fig.\ \ref{fig:Fig1}.}

\hili{Comparison with DFT is repeated for one of the peptide not included in the training set. The polarizability with time and Raman spectra of GAG are presented in Fig.\ \ref{fig:Fig3}(c) and (d), respectively. Note that for GAG, trajectories of 6.5 ps are used to obtain Raman spectra. Although this represents very short trajectories and definitely does not sample all possible conformations, it can still be used to compare the different models with DFT results. Similarly to glycine, polarizabilites predicted by ML models is in better agreement with DFT than the Thole model, Fig.\ \ref{fig:Fig3}(c). Additionally, TNEP seems more accurate than SA-GPR. Fig.\ \ref{fig:Fig3}(d) then compares Raman spectra of GAG from the different models with DFT. Spectra from SA-GPR and the Thole model show discrepencies with DFT results, while TNEP is found to be in much better agreement.}
\hili{\remove{Overall, TNEP is found to show the best agreement with DFT spectra. Also note that polarizabilities for peptides not included in the training step are better predicted by TNEP [see Fig.\ \ref{fig:Fig2}(b-e)].}} This latter model is therefore used to evaluate the spectra in the rest of this work. 
We also show comparison to the results from Thole model to evaluate its accuracy for simulating Raman spectra.

\begin{figure*}[t] 
    \centering
    \includegraphics[width=7in]{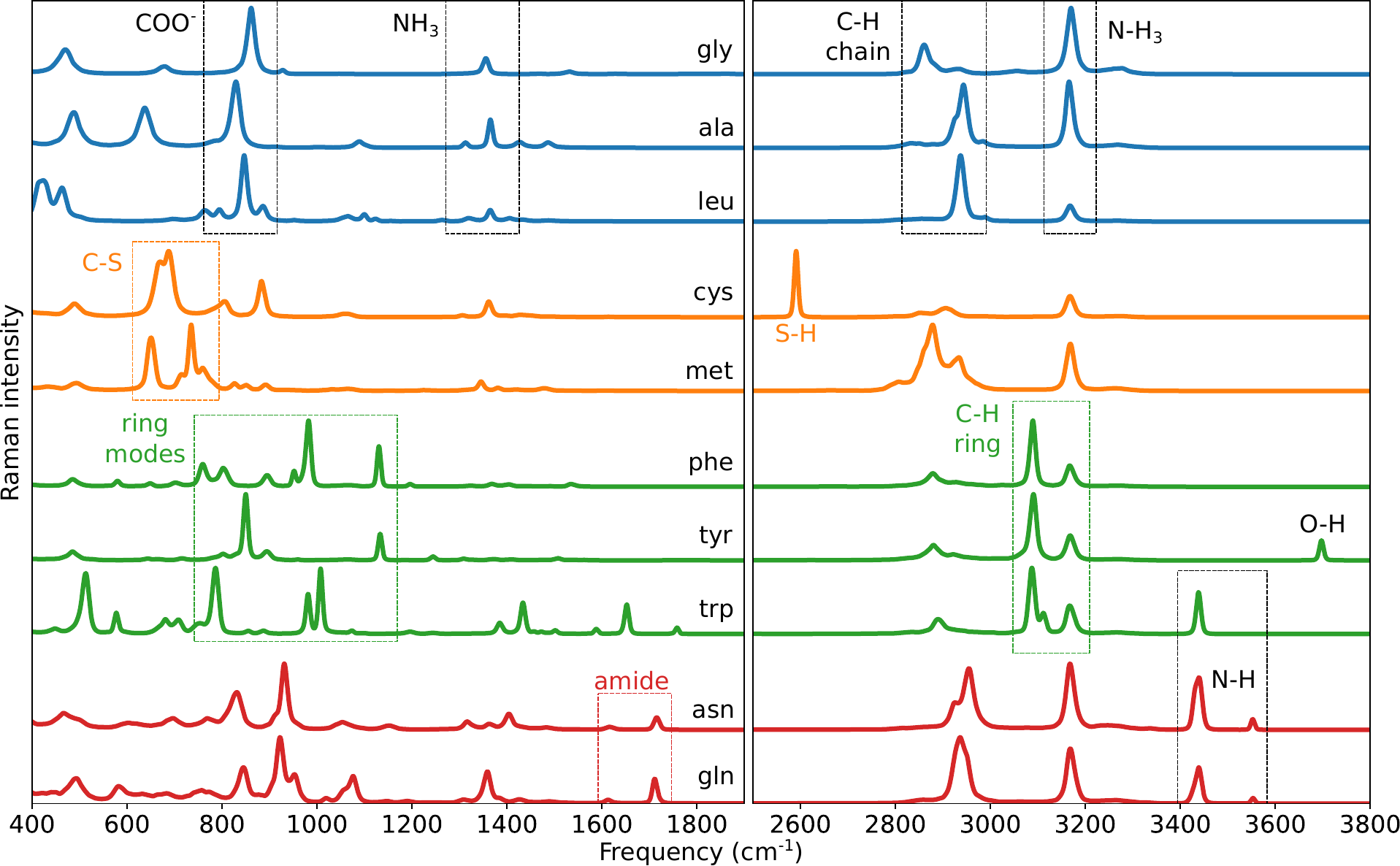}
    \caption{Low- and high-frequency region of the Raman spectra for some amino acids in aqueous solution. A complete figure containing all 20 amino acids can be found in the SI. Labels show peak assignment, where black labels correspond to general mode and colored labels refer to modes specific to the side chain: hydrocarbons in blue, sulfur in orange, aromatic in green, and amide in red.}
    \label{fig:Fig5}
\end{figure*}

The quality of TNEP spectra are further compared to experimental measurements from the database presented in Ref.\ \citenum{DeGelder2007}. Results are shown in Fig.\ \ref{fig:Fig4} for glycine, alanine, phenylalanine, serine and arginine. TNEP spectra are very similar to the ones from experiments for every amino acids, with most peak being correctly reproduced (peak attributions is discussed later in Fig.\ \ref{fig:Fig5}). Even though spectra are normalized, relative intensities between peaks are also in good agreement. \hili{Peak intensities greatly depend on the quality of the polarizabilities calculations/predictions, further confirming the good training of TNEP. On the other hand,} peak frequencies show larger discrepancies. For gly, ala and phe, most intense peaks are shifted by 20 to 30 \icm. Larger shifts are observed in the spectrum of serine, with differences up to 50 \icm. While the discrepancies could originate from many sources, including particularities of the experimental measurements, \hili{differences in the peak position can usually be attributed to the quality of the MD trajectories and the choice of force field. In order to get an idea of the impact of the force field on Raman spectra, we also performed MD simulations using OPLS force field. A comparison between CHARMM and OPLS can be found in Fig.\ S6 in the SI, and clearly shows larger spectral differences than between the polarizability models. One difference between OPLS and CHARMM is the rotation of the C-terminus, which is present in the former but not the latter. Such rotation also happens when using other force fields \cite{Horn2014}. 
\remove{A careful comparison between force fields could further improve the quality of spectra and lead to better agreement with experiments.}
In the case of amino acids and peptides, errors might also arise from incomplete sampling of the various conformations. This could also explain in part the differences with experimental results. Additionally, our calculations only include the isotropic part of the Raman spectra, which could explain why some of the low intensities peaks are missing (typically in glycine around 1100 \icm). We also note that TNEP shows great agreement with experimental observations for arginine, although the simulation of this amino acid is challenging due to its charge and its long side chain (leading to different conformations).}

Raman spectra of amino acids can be separated in two regions. The region below 1800 \icm\ contains peaks due to the movement of carbon, oxygen and nitrogen, including aromatic modes around 1000 \icm\ and very-low frequency modes due to the change of conformations. Hydrogen modes are found at higher frequencies between 2600 and 3800 \icm. Raman spectra in both regions are presented in Fig.\ \ref{fig:Fig5}.
The low-frequency region between 100 and 1800 \icm\ contains most Raman peaks of amino acids. These peak can be used to identify amino acids in peptides \cite{Candeloro2013}, even though peak attribution can be quite challenging due to their high number. Fig.\ \ref{fig:Fig5} presents Raman spectra of only a few amino acids to facilitate the discussion, but results for all 20 amino acids can be found in the SI (see Fig.\ S7). 
The Raman spectrum of glycine shows two sharp peaks at 860 and 1350 \icm, which agree well with experimental measurements \cite{Krauklis2020}. These peaks are attributed to the \ce{COO-} and the amine group, respectively \cite{Derbel2007}. 
For other amino acids with hydrocarbon side chains, such as alanine and leucine, the same two peaks are observed (also see Fig.\ S3). 
Note that the \ce{COO-} peak shifts to 830 \icm\ for alanine, in good agreement with experimental measurements \cite{Hernandez2009,Zhu2011}. Longer side chains also lead to additional peaks, with for example peaks at 1430 and 1490 \icm\ for alanine which correspond to rocking of the \ce{-CH3} chain. Same \ce{COO-} and amine peaks are observed for cysteine. In this case, an additional peak due to the \ce{C-S} bond appears at 680 \icm, in good agreement with experiments \cite{Hernandez2011,Swiech2023}. Methionine, the other amino acid with a sulfur side chain, possesses similar peaks between 650-720 \icm \cite{Jenkins2005,Candeloro2013,Hernandez2011}. In Fig.\ \ref{fig:Fig5} one can see such two peaks at 650 and 730 \icm\ which can be attributed to the two different \ce{C-S} bonds. For aromatic side chains, high intensity is observed for the breathing mode of benzene at 990 \icm, which is correctly reproduced for phenylalanine. This peak shifts to 850 \icm\ in tyrosine due to the additional \ce{OH} added to the ring. For these two amino acids, another intense peak is found at 1130 \icm\ and correspond to another mode within the benzene ring \cite{Rava1985} (denoted 9a in Ref.\ \citenum{Wilson1934}). For tryptophan, which contains both a pyrrole and a benzene ring, breathing modes are observed at 790 and around 1000 \icm\ respectively, in good agreement with experiments \cite{Rava1985}. Asparagine and glutamine, two amino acids with amide side chains, possess similar Raman spectra. In particular, they both show an unique peak at 1710 \icm, which has been observed experimentally and assigned to the amide \cite{Baran2007,Sylvestre2014}.

\begin{figure*}[t] 
    \centering
    \includegraphics[width=7in]{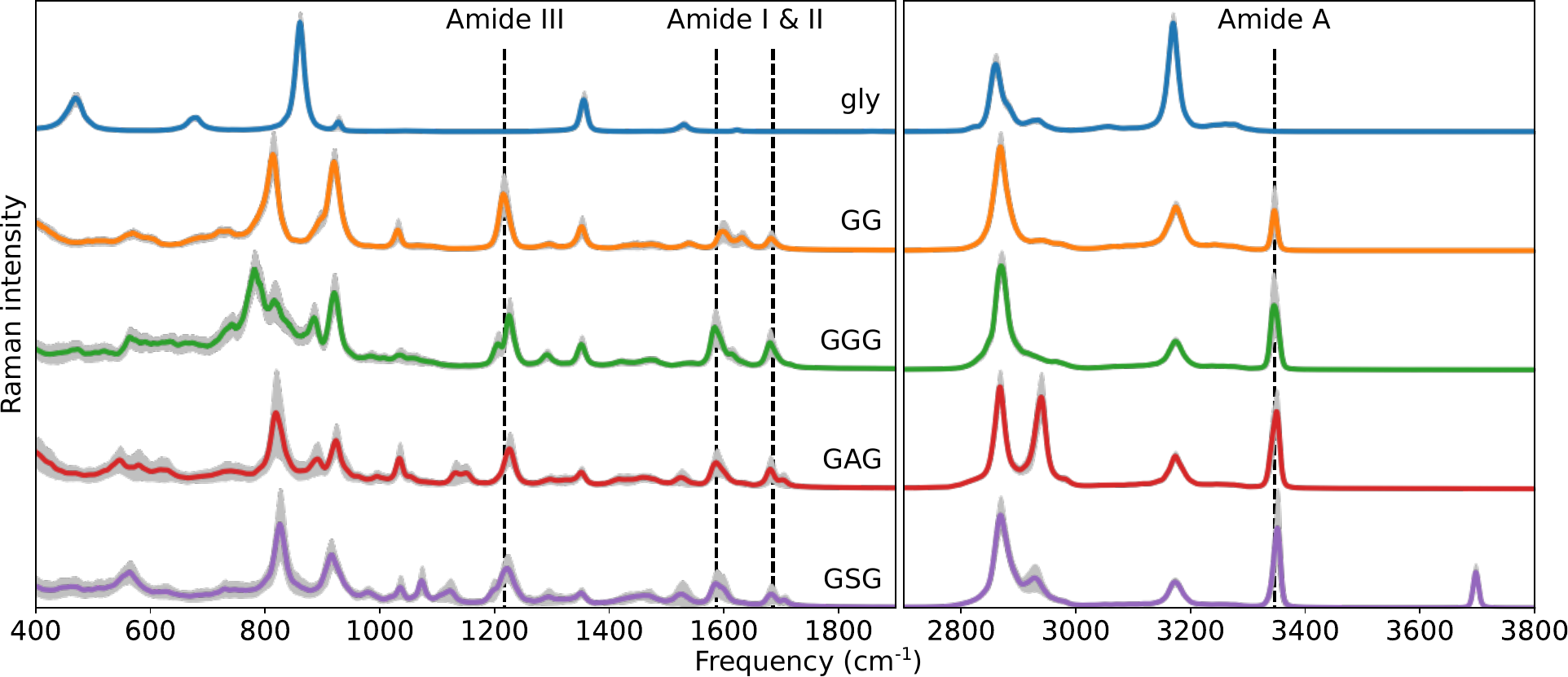}
    \caption{Raman spectra of glycine and glycine-based peptides. Left/right panels show the low/high-frequency regions. Committee error estimates are shown using shaded grey area. Selected peaks labels are also indicated.}
    \label{fig:Fig6}
\end{figure*}

The high-frequency region of the Raman spectra is also shown in Fig.\ \ref{fig:Fig5}. Cysteine is the only molecule showing a very clear peak around 2600 \icm, which can be attributed to the S-H bond. Broad peaks between 2800-3100 \icm\ are due to the C-H vibrations. For amino acids with aromatic side chains, sharp peaks around 3100 \icm\ can be attributed to C-H modes inside the carbon ring. Most amino acids show a clear peak around 3200 \icm\ which correspond to the hydrogen motion inside the amino group (\ce{NH3}). Proline is the only amino acid without such group and does not show this peak (see Fig.\ S7 in the SI). Instead, a peak at higher frequency is observed and attributed to the N-H bond. Other peaks between 3400 and 3600 \icm\ are found for tryptophan, asparagine and glutamine, which are also attributed to N-H bonds. Finally, O-H bonds which are present in tyrosine lead to one peak around 3700 \icm. This peak attribution is in great agreement with previous experimental results \cite{Jenkins2005}. Overall, Raman spectra obtained by combining CHARMM27 trajectories and polarizabilities from TNEP correctly reproduce experimental spectra for all kinds of side chains and correctly account for hydrogen motion, sulfur vibrations, aromatic rings and amide groups.

\subsection{Raman spectra of peptides and error estimates}

\begin{figure*}[t] 
    \centering
    \includegraphics[width=7in]{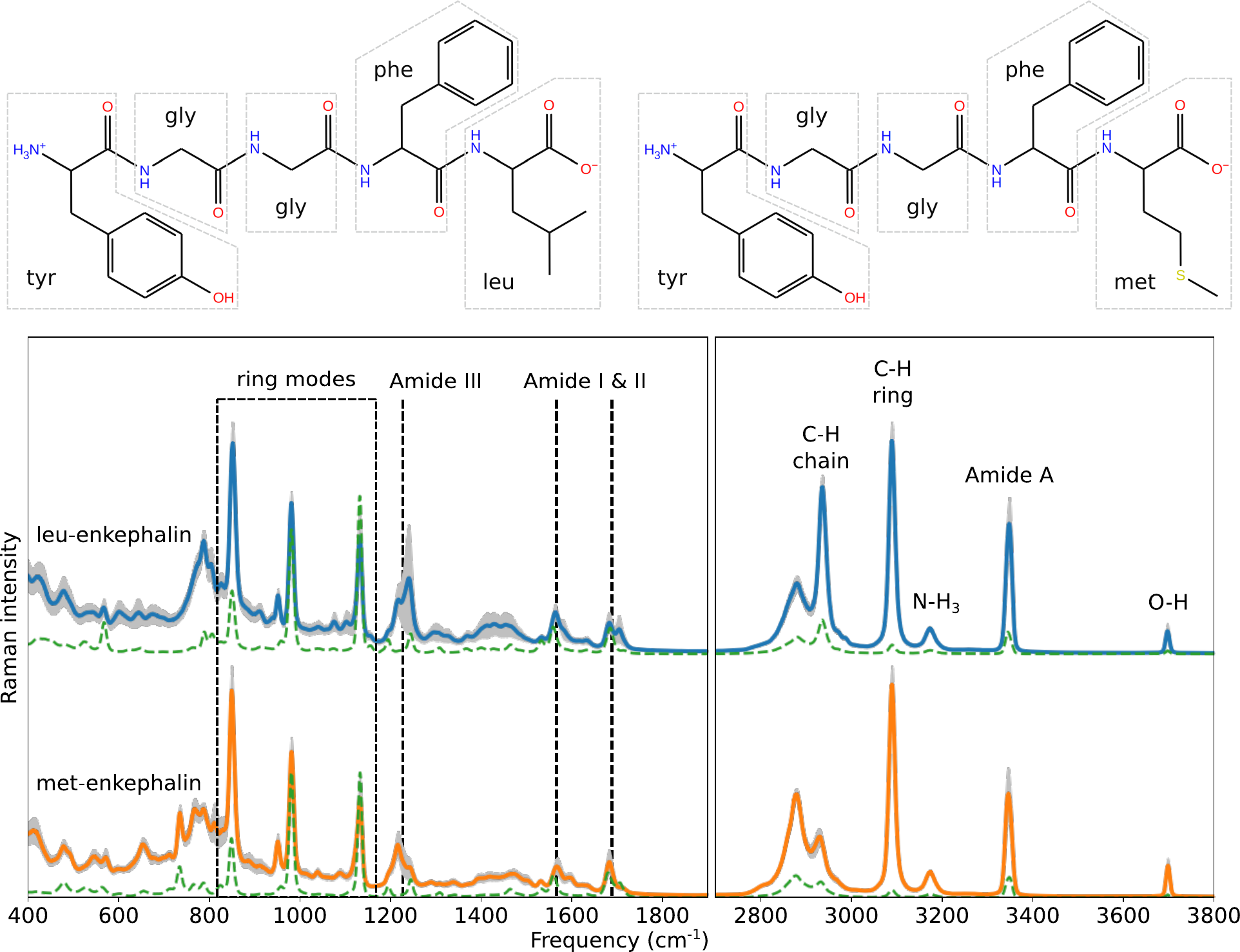}
    \caption{Top: representation of leu-enkephalin (left) and met-enkephalin (right). Bottom: Raman spectra of leu- and met-enkephalin. Committee error estimates are represented with grey area and spectra from the Thole model (multiplied by a factor 5000) in dashed green. Peaks are labeled using the results presented in Fig.\ \ref{fig:Fig5} and \ref{fig:Fig6}.}
    \label{fig:Fig7}
\end{figure*}

When simulating the spectra, especially of larger peptides that were not explicitly included in the training set, it is important to be able to estimate the quality of the predictions.
Here we start with glycine-based peptides included in the training set, as shown in Fig.\ \ref{fig:Fig6}. Committee error estimates averaged over 10 committee models are used to \hili{estimate} errors on spectra, which are represented by shaded grey areas. 
For single glycine, errors are negligible for all features, showing that the models are very well trained for this molecule. For di-glycine (GG), errors are low for most peaks, with largest errors appearing between 1600-1700 \icm. These regions contains two peaks attributed to the amide formed by the peptides bonds and are usually labeled as amide I and II \cite{Hayashi2007,Yan2020}. These peaks are particularly visible for GGG at 1580 and 1680 \icm\ and similar peaks in the same region were already observed for amino acids containing amide, namely asparagine and glutamine (see Fig.\ \ref{fig:Fig5}). Large errors in this region can therefore be attributed to the relatively small training done on peptides and peptide bonds. Another mode labeled amide III usually appears between 1200-1300 \icm\ \cite{Hayashi2007,Yan2020}. Such a peak is clearly observed in our results between 1210 and 1230 \icm\ depending on the peptide. 
On the other hand, peaks previously observed in single glycine change due to peptide bonding. The peak at 860 \icm\ attributed to the \ce{COO-} group is replaced by two distinct peaks pertaining to the C-O and C-N vibrations in peptide bonds. The C-O peak also shifts due to the \ce{COO-} environment being replaced by N-C-O. Additionally, the \ce{NH3} group present in single amino acids disappears during peptide bonding and only one terminal \ce{NH3} remains in peptides, leading to a decrease of intensity for the peak at 1350 \icm. 
Since they are not included in the training set, GAG and GSG spectra show larger errors, even though they remain reasonable. Note that for these two peptides, largest errors are observed at very low frequencies corresponding to conformational changes. While bond vibrations of peptides can be reasonably learned from amino acids/other peptides, conformational changes are unique to each peptides and cannot be directly transferred.

In the high-frequency region, errors are much smaller, probably due to the hydrogen motion being less dependent on their surrounding environments. Another new amide peak usually labeled amide A is observed at 3350 \icm\ for all peptides \cite{Hayashi2007,Yan2020}. The intensity of previously observed peaks around 3200 \icm\ pertaining to the \ce{NH3} hydrogen motion decreases for peptides. Due to the smaller training done on amide A, this new mode exhibits the largest error in this frequency range. One can also note the peak observed at 3700 \icm\ in GSG due to the \ce{OH} bond in serine. Features of Raman spectra of peptides are correctly reproduced in both low- and high-frequency regions. In particular, peaks attributed to \ce{COO} and \ce{NH3} disappear in peptides and new peaks from amide can be observed instead. Overall, positions of these new amide peaks in both frequency regions agree well with previous calculations and experiments on peptides  \cite{Hayashi2007,Yan2020,Stewart1999}, as well as results on amides in other molecules \cite{Chen1995,Herrebout2001,Asakura2021}. \hili{Similar spectra are also obtained when using SA-GPR polarizabilities (see Fig.\ S8 in the SI).}

Finally, we investigate Raman spectra of larger pentapeptides not included in the training step, namely met- and leu-enkephalins. These opioid peptides possess analgesic properties similar to morphin and play an important role in stress resilience \cite{Hughes1975,Takeuchi1992,Henry2017}. They take the form tyr-gly-gly-phe-met and tyr-gly-gly-phe-leu for met- and leu-enkephalin, respectively, as illustrated in Fig.\ \ref{fig:Fig7}, along with resulting spectra and their errors. Raman spectra are fairly similar for both of these peptides. Like other peptides, large errors are observed in the very low frequencies region below 400 \icm\ related to conformational changes. Low-frequency region is mainly dominated by peaks pertaining to the aromatic rings, with peaks at 850 \icm, 980 \icm\ and 1130 \icm, which respectively correspond to the breathing mode of tyrosine and phenylalanine and the 9a mode already observed for isolated amino acids (see discussion on Fig.\ \ref{fig:Fig5}). Errors for these three modes are relatively small. Additionally, amide III peak is also visible at 1230 \icm\ as well as amide II and I at 1560 and 1680 \icm, in good agreement with previous results from smaller peptides (see Fig.\ \ref{fig:Fig6}). Note that here again, amide peaks exhibit larger errors than modes coming from amino acids. In the high-frequency region, observed peaks are in great agreement with what was previously obtained for amino acids (Fig.\ \ref{fig:Fig5}) and peptides (Fig.\ \ref{fig:Fig6}). In this regions, peaks from aromatic rings and amide are the most intense. Also note that modes from \ce{CH} vibrations show a wide peaks due to the many different environments existing in such large peptides. Peaks from \ce{NH3} and \ce{OH} are also observed with smaller intensities. Committee errors are also much smaller in this frequency region. 

Raman spectra from the Thole model are also shown in in Fig.\ \ref{fig:Fig7} to assess its accuracy. Most of the peaks are correctly picked out, including both ring breathing modes, the 9a mode at 1130 \icm\, and the amide I and II modes. In the higher frequencies region, all modes compare well, even though the Thole model clearly underestimates all intensities (similar to what was previously observed for glycine in Fig.\ \ref{fig:Fig3}). \hili{Similar to glycine-based peptides, SA-GPR also correctly reproduce most peaks for enkephalins, as can be seen from Fig.\ S8 in the SI.}

Overall, based on the fairly good agreement with experiments\cite{Niederhafner2021} and the overall low committee error estimates, the Raman spectra of enkephalins should be fairly accurate even though these peptides were not included in the training. Moreover, committee error estimates can be useful for assessing which peaks are likely to be less accurate and guide possible further training. In the case of enkephalins, amide and conformational changes could benefit from further training. 

\section{Conclusion}

We studied the accuracy and transferability of machine learning models in predicting polarizabilities of amino acids and small peptides. Two models based on TNEP and SA-GPR were trained and compared, and while SA-GPR is more accurate for single molecule predictions, we find that TNEP (and probably neural network in general) are more suited for a large set of molecules such as amino acids. Additionally, we find that addition of peptide bonds to the training set is important in correctly predicting polarizabilities of peptides. Models trained on all amino acids and simple peptides can also accurately predict polarizabilities for peptides not included in the training set.

Polarizability models were then combined with classical MD to obtain the Raman spectra. Resulting spectra are in fairly good agreement with experimental observations and correctly reproduce many of the different modes that can be observed in all amino acids. 
The errors in peak frequencies arise from the classical force field used in MD and thus further improvement of the spectra requires development of more accurate (machine-learning) force field. 
\hili{With the models developed herein, it becomes possible to benchmark the force field against experimentally measured spectra.}
For peptides, committee error estimates show that spectral features from modes that are less trained (typically conformational changes and amide vibrations) are generally less accurate. Raman spectra of enkephalin are also investigated and show good agreement with experiment, even though polarizability models were not trained on these larger peptides.

The polarizability model developed in this work could be immediately used in future studies on peptides.
Even though the accuracy of the model for much larger molecules is still unknown, it could be used as a very good starting point and would only need to be retrained for the different conformations to be applicable to a particular protein.

\section*{Data and software availability statement}

GROMACS 2022.4, cp2k 2023.1 and GPUMD 3.7 are used and are open source software available on their respective websites. SA-GPR models are trained using the TENSOAP package available at https://github.com/dilkins/TENSOAP. Methodology for the production of MD simulations are presented in the Computational details section of the main test. GROMACS, cp2k and GPUMD input files, polarizability training/test sets and resulting TNEP models are available on a public repository at DOI:10.5281/zenodo.10491770.

\section*{Supporting information}

\hili{Effect of solvation on the polarizability, Transferability at larger cutoffs,} benchmark/training of the committee error estimates models\hili{, Raman spectra from different training set size, Comparison between force fields, }Raman spectra of all amino acids \hili{and Raman spectra using SA-GPR.}

\section*{Acknowledgment}

We acknowledge funding from Research Council of Finland (project No. 357483). We also thank CSC–IT Center for Science Ltd. for generous grants of computer time. We acknowledge Biocenter Finland for provided services.

\bibliography{raman_BAs}

\end{document}


\author{Ethan Berger}
\affiliation{Microelectronics Research Unit, Faculty of Information Technology and Electrical Engineering, University of Oulu, P.O. Box 4500, Oulu, FIN-90014, Finland}
\author{Juha Niemelä}
\affiliation{Faculty of Biochemistry and Molecular Medicine, University of Oulu, Oulu, Finland}
\author{Outi Lampela}
\affiliation{Faculty of Biochemistry and Molecular Medicine, University of Oulu, Oulu, Finland}
\author{Andre Juffer}
\affiliation{Faculty of Biochemistry and Molecular Medicine, University of Oulu, Oulu, Finland}
\author{Hannu-Pekka Komsa}
\affiliation{Microelectronics Research Unit, Faculty of Information Technology and Electrical Engineering, University of Oulu, P.O. Box 4500, Oulu, FIN-90014, Finland}
\email{hannu-pekka.komsa@oulu.fi}

\title{Supporting Information: \\ Raman spectra of amino acids and peptides from machine learning polarizabilities}
\maketitle

\section{Effect of solvation on the polarizability}

\begin{figure}[h]
    \centering
    \includegraphics[width=0.5\linewidth]{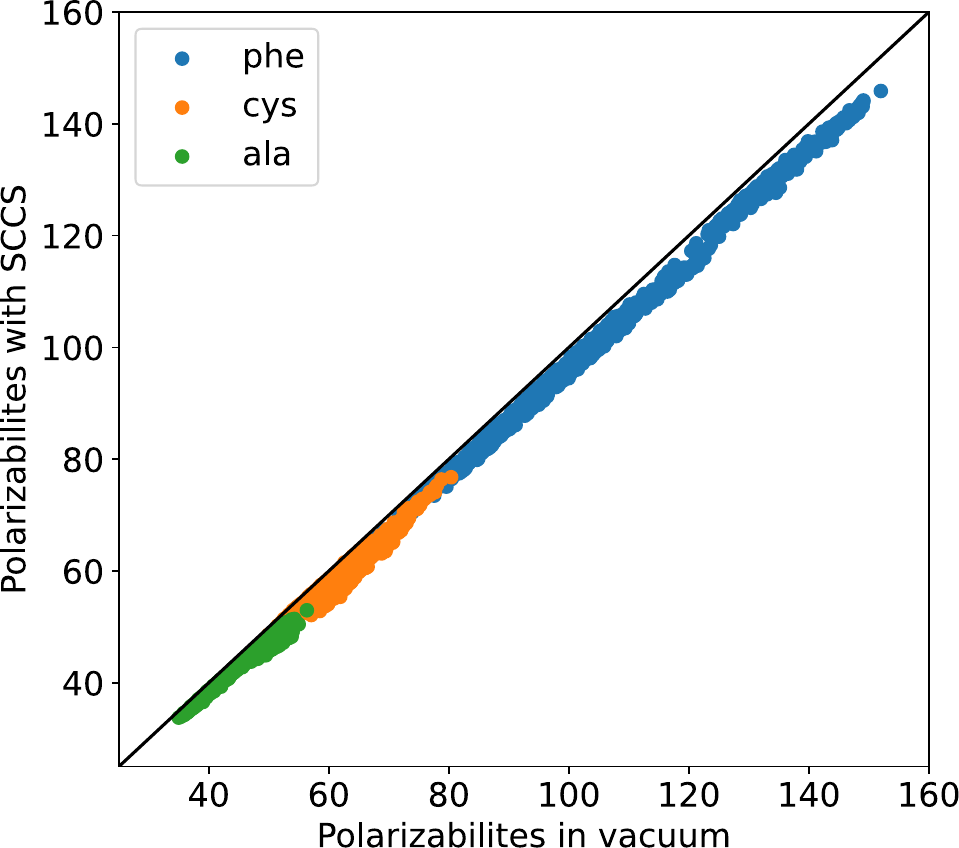}
    \caption{Comparison between polarizabilities from DFT calculations in vacuum and using SCCS.}
    \label{fig:FigS1}
\end{figure}

Explicitly including water molecules as solvent in DFT calculation would prove computationally impossible. Accounting for the effect of solvation in DFT calculation is therefore a difficult task and simply considering molecules in vacuum would be much easier. To this end, we briefly investigated the effect of solvation on the polarizabilities of a few amino acids. IN particular we use the self-consistent continuum solvation method as implemented in cp2k. Results are presented in Fig.\ \ref{fig:FigS1}. It is pretty clear that the effect of solvation are very small, only showing a small constant shift. Such a shift will not impact the resulting Raman spectra and we therefore decide to treat amino acids and peptides in vacuum when performing DFT calculations. Note that the effect of solvation is still explicitly included during classical force field MD.

\pagebreak

\section{Transferability at larger cutoffs}

The choice of cutoff distance has a fairly large impact on the transferability of each models. For GG and GGG which are included in the training, both TNEP and SA-GPR lead to similar accuracy for all cutoffs. For other peptides, smaller cutoffs are found to be beneficial in general, with minimal RMSE at 4 \AA\ and 6 \AA\ for TNEP and SA-GPR respectively.

\begin{figure*}[h!] 
    \centering
    \includegraphics[width=\linewidth]{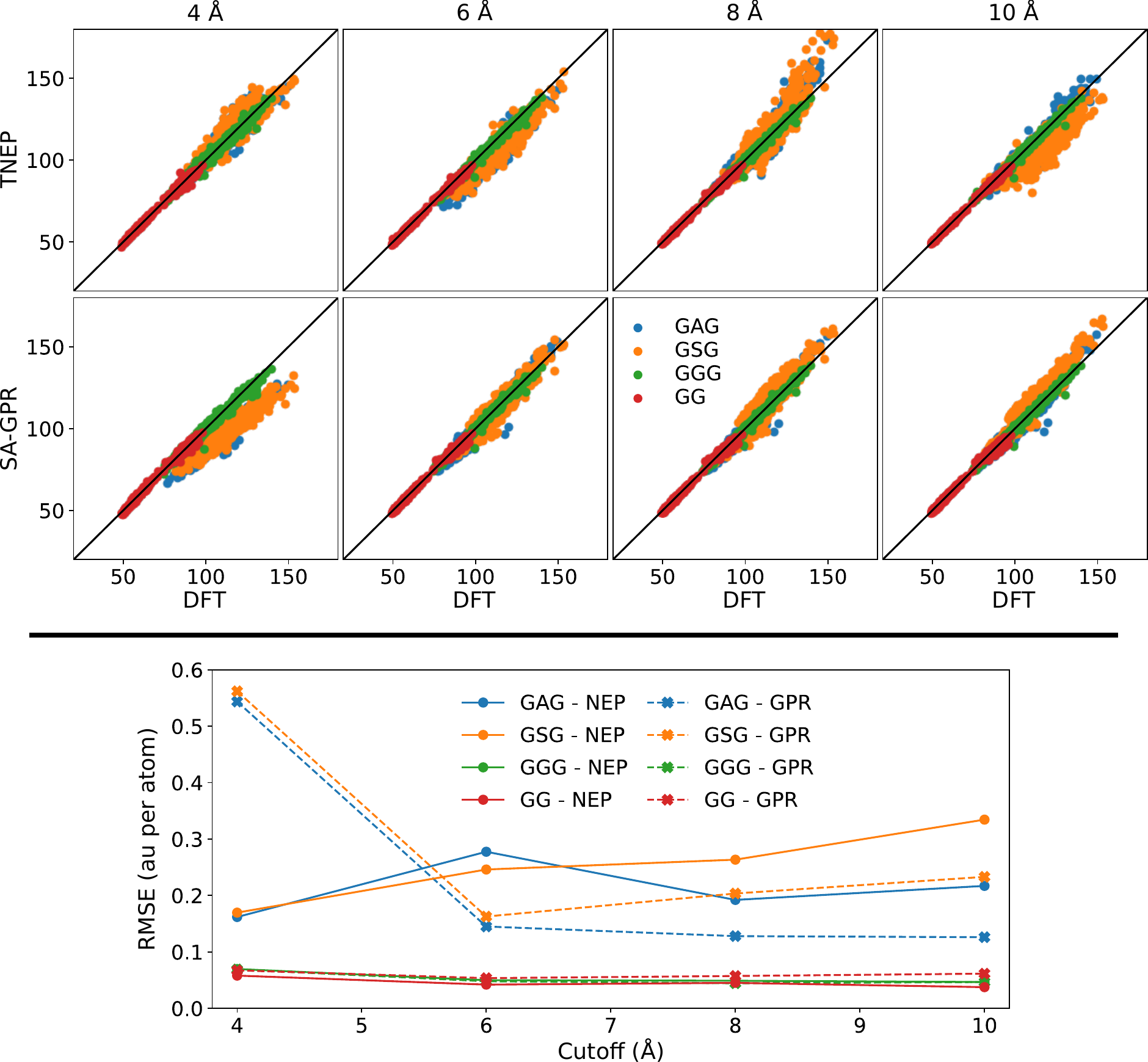}
    \caption{Top : Comparison between polarizabilites from DFT and the ML models for glycine-based amino acids. Each panel show results for different cutoff distances. Bottom : RMSE of each peptides with increasing cutoff distances.}
    \label{fig:FigS2}
\end{figure*}

\pagebreak

\section{Benchmark/training of the committee error estimates models}
\begin{figure*}[h!] 
    \centering
    \includegraphics[width=\linewidth]{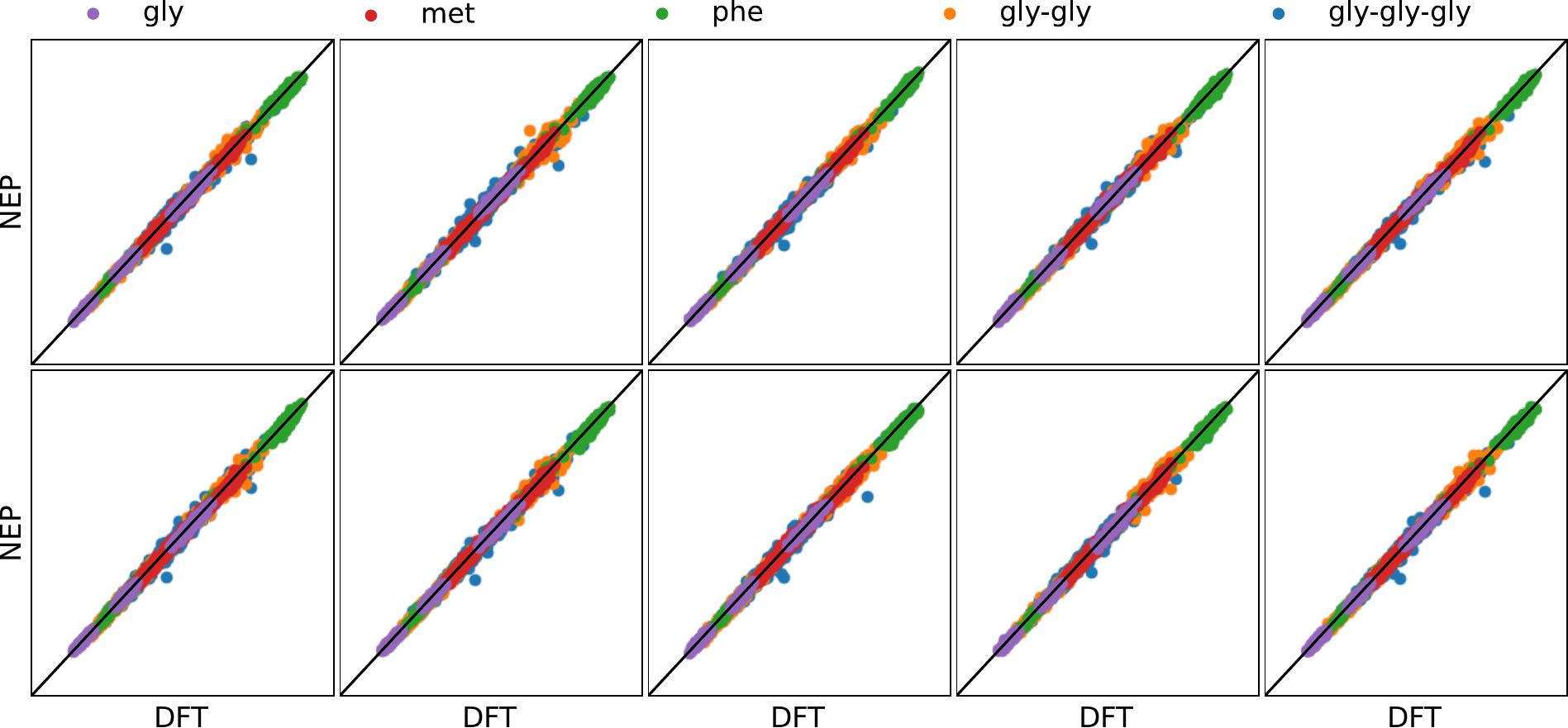}
    \caption{Comparison between polarizabilities from DFT calculations and NEP predictions for a few amino acids and peptides. Each panels represent one of the committee model.}
    \label{fig:FigS3}
\end{figure*}

Committee error estimate is used to asses the quality of the NEP model for large peptides. To this end, 10 NEP models are trained using the same training set. Using the fact that the SNES optimization algorithm is random, this results in 10 different neural networks. Fig.\ \ref{fig:FigS3} shows the comparison between DFT calculations and NEP prediction for each of these 10 models. It is pretty clear that all models are well trained and lead to similar accuracy.

\pagebreak

\begin{figure*}[h!] 
    \centering
    \includegraphics[width=\linewidth]{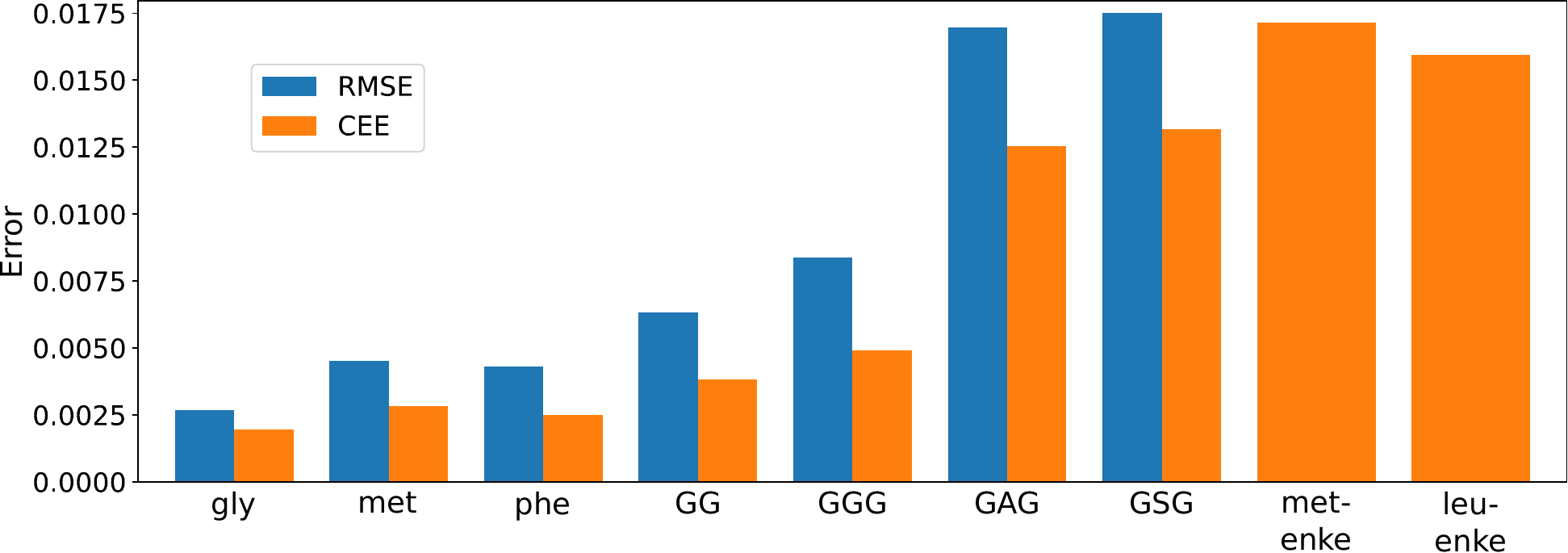}
    \caption{Comparison between root mean squared error (RMSE) and committee error estimates (CEE) for a few amino acids and peptides.}
    \label{fig:FigS4}
\end{figure*}

The resulting committee error estimates are then benchmarked and compared to the real error (RMSE). Results are shown in Fig.\ \ref{fig:FigS4}. Note that errors per atoms are used. Errors are small for single amino acids and increase for peptides. For GG and GGG, which are included in the training set, errors remain below 0.01, while they are larger for GAG and GSG which are not included in the training set. For every molecules, CEE underestimates the RMSE but overall follows the same trend and agree well qualitatively. For larger peptides (met- and leu-enkephalins), there are no DFT calculations and hence no RMSE. Instead, only CEE is shown and it shows only a small increases compared to GAG and GSG. Also note that met-enkephalins shows a slightly larger CEE, probably due to the sulfur atom being less well trained (mostly because only two amino acids contain one sulfur atom, leading to only little training for this atom).

\pagebreak

\section{Raman spectra from different training set size}

\begin{figure*}[h!] 
    \centering
    \includegraphics[width=\linewidth]{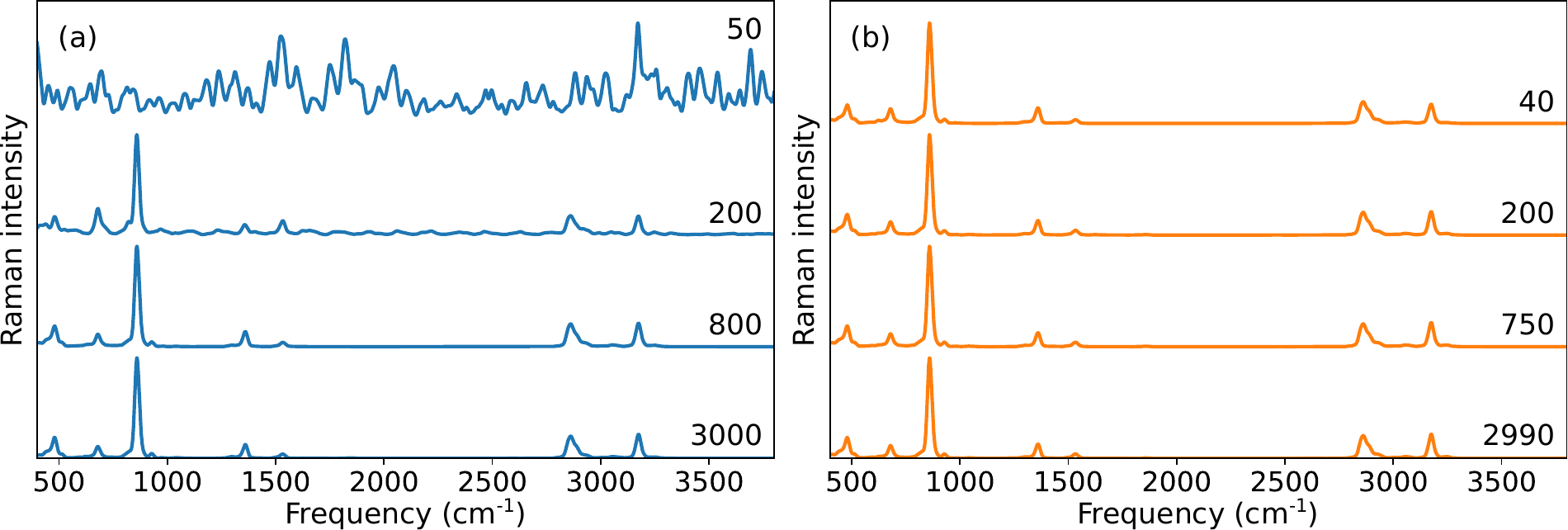}
    \caption{Raman spectra of glycine from different training set sizes using (a) NEP and (b) SA-GPR.}
    \label{fig:FigS5}
\end{figure*}

Knowing how accurate polarizability models have to be to correctly predict Raman spectra is a central question. Fig.\ \ref{fig:FigS5} shows Raman spectra from models trained using different amount of data. For NEP, showed in panel (a), we find that using 50 structures for the training lead to completely wrong spectra. Increasing this number to 200 lead to better yet still very noisy spectra. High quality is achieved only when using 800 structures or more. From Fig.\ 1 in the main text, this correspond to a RMSE of $~10^{-2}$ au per atoms. On the other hand, SA-GPR, showed in panel (b), lead to correct spectra even for training sets containing only 40 structures. This further support the fact that SA-GPR performs better for small amount of training data.

\pagebreak

\section{Comparison between force fields}

\begin{figure*}[h!]
    \centering
    \includegraphics[width=\linewidth]{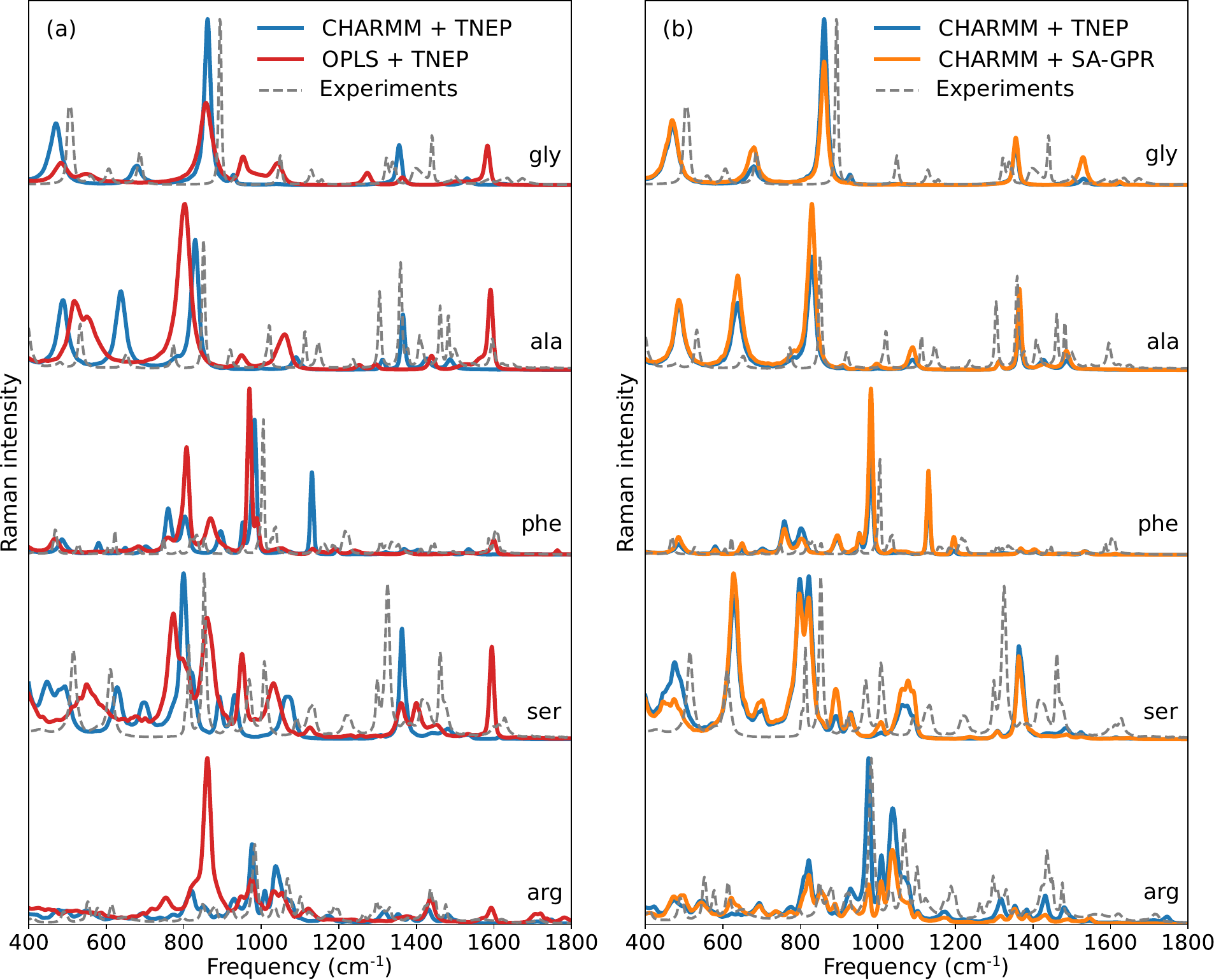}
    \caption{Comparison between (a) force fields and (b) polarizability models. Experimental results are shown with grey lines.}
    \label{fig:FigS6}
\end{figure*}

Fig.\ \ref{fig:FigS6} (a) shows a comparison between CHARMM 27 and OPLS while Fig.\ \ref{fig:FigS6}(b) compares TNEP and SA-GPR. Experimental results are also shown in grey. Note that TNEP was not retrained for OPLS trajectories. Contrary to other figures, simulated spectra are not normalized here for a better comparison between methods. Experimental spectra are normalized to the CHARMM+TNEP spectra. We find that polarizability model mainly impacts the intensity of the peaks, while force fields also modifies position and widths.

\pagebreak

\section{Raman spectra of all amino acids}

\begin{figure*}[h!] 
    \centering
    \includegraphics[width=\linewidth]{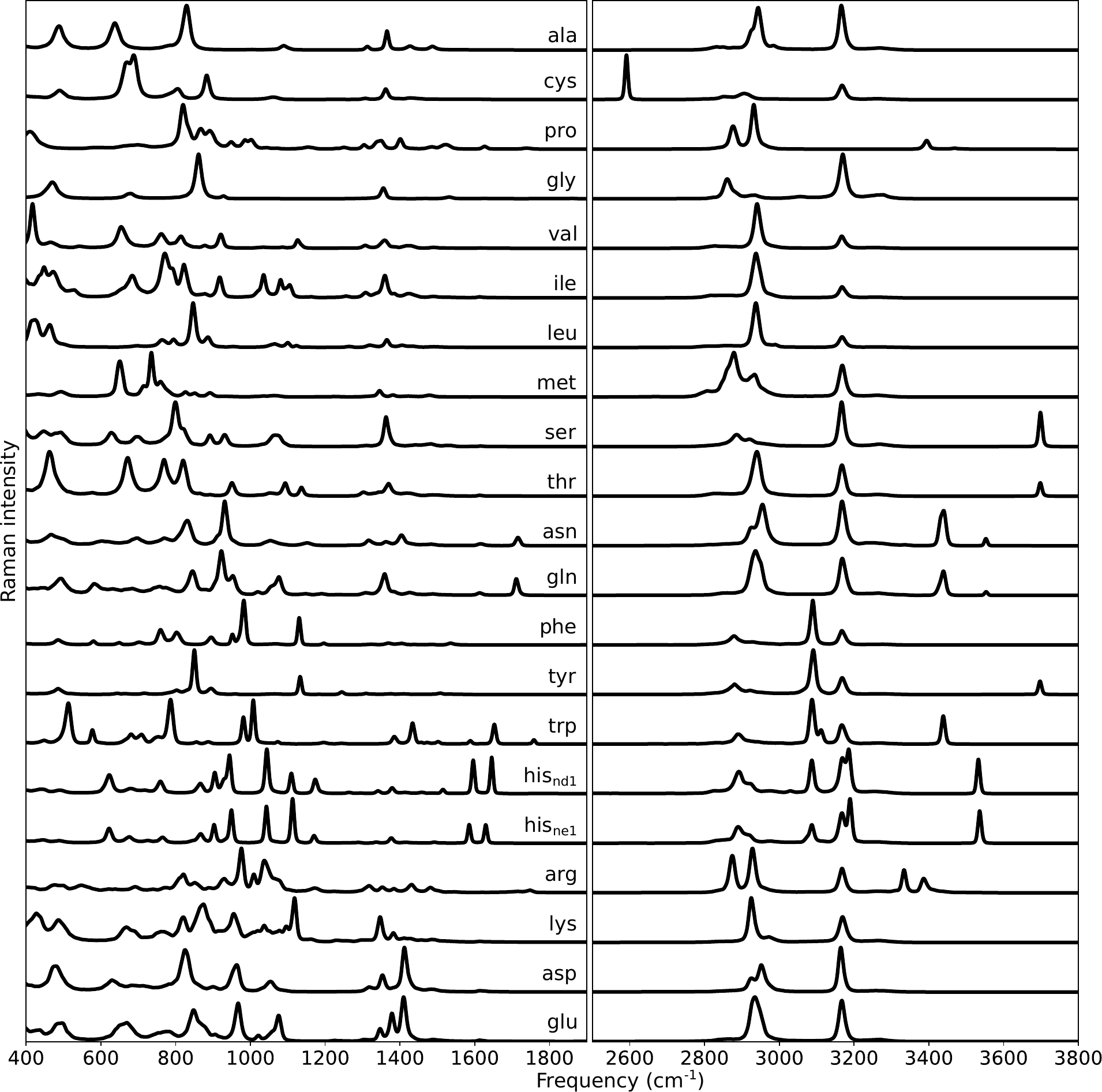}
    \caption{Raman spectra of all 20 amino acids. Left and right panels show the low- and high-frequency regions, respectively.}
    \label{fig:FigS7}
\end{figure*}

While only Raman spectra of only a few selected amino acids are discussed in the main text, Fig.\ \ref{fig:FigS7} shows Raman spectra for all 20 amino acids. Note that two different form of histidine are shown, eventhough they lead to the same spectra.

\pagebreak

\section{Raman spectra using SA-GPR}

\begin{figure*}[h!] 
    \centering
    \includegraphics[width=\linewidth]{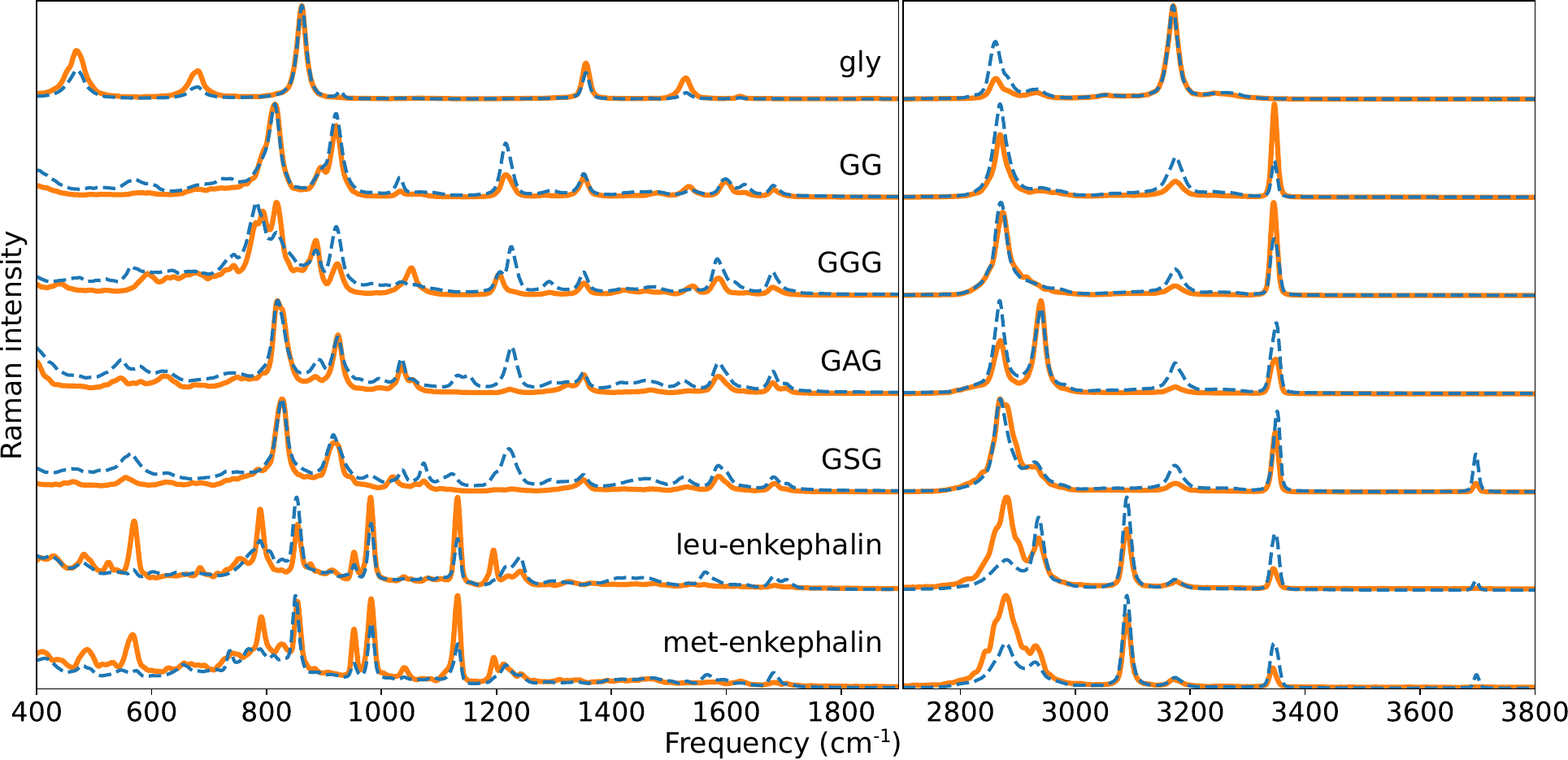}
    \caption{Raman spectra of glycine peptides and enkephalins. Solid orange lines show spectra obtained from SA-GPR polarizabilites while dashed blue line come from TNEP.}
    \label{fig:FigS8}
\end{figure*}

Fig.\ \ref{fig:FigS8} compares Raman spectra of peptides using polarizabilities from TNEP and SA-GPR. From the results in Fig.\ \ref{fig:FigS2}, cutoffs of 4 \AA\ and 6 \AA\ are used for TNEP and SA-GPR, respectively. Due to the higher computational cost of SA-GPR, spectra are obtained using only the first nanosecond of the production trajectories. Also note that shorter trajectories are for SA-GPR, which can lead to small discrepancies with TNEP. 

Both polarizability method lead to qualitatively similar spectra, although the intensity for some peaks differ. In particular, large discrepancies are found for amide III. In the main text, larger errors are found for this peaks using committee error estimates and are attributed to limited training of peptides bonds, which could also explain part of the differences observed here.

\pagebreak